\documentclass[sigconf]{acmart}
\AtBeginDocument{%
  }

\copyrightyear{2026}
\acmYear{2026}
\setcopyright{cc}
\setcctype{by-nc-nd}
\acmConference[KDD '26]{Proceedings of the 32nd ACM SIGKDD Conference on Knowledge Discovery and Data Mining V.2}{August 09--13, 2026}{Jeju Island, Republic of Korea}
\acmBooktitle{Proceedings of the 32nd ACM SIGKDD Conference on Knowledge Discovery and Data Mining V.2 (KDD '26), August 09--13, 2026, Jeju Island, Republic of Korea}
\acmDOI{10.1145/3770855.3818501}
\acmISBN{979-8-4007-2259-2/2026/08}

\usepackage{graphicx}
\usepackage{subcaption}

\begin{document}

\title{Uncertainty-Aware Adaptive Recommendation across User Lifecycle}
\author{Bob Junyi Zou}
\email{junyizou@stanford.edu} 
\affiliation{%
  \institution{Stanford University}
  \city{Stanford}
  \state{CA}
  \country{USA}
}

\author{Sai Li}
\email{zhuning.1@bytedance.com}
\affiliation{%
  \institution{ByteDance Inc.}
    \city{San Jose}
  \state{CA}
  \country{USA}
}

\author{Tianyun Sun}
\email{tianyun.sun@bytedance.com}
\affiliation{%
  \institution{ByteDance Inc.}
    \city{San Jose}
  \state{CA}
  \country{USA}
}

\author{Wentao Guo}
\email{wentao.guo@bytedance.com}
\affiliation{%
  \institution{ByteDance Inc.}
    \city{San Jose}
  \state{CA}
  \country{USA}
}

\author{Qinglei Wang}
\email{wangqinglei@bytedance.com}
\affiliation{%
  \institution{ByteDance Inc.}
    \city{Beijing}
  \state{Beijing}
  \country{China}
}


\begin{abstract}
A fundamental challenge in recommender systems is balancing reliability for Low-Active Users (LAUs) with diversity for High-Active Users (HAUs). The key to this balance lies in quantifying model uncertainty, which approximates the risk of prediction errors and reveals the limits of the model's current knowledge. On large-scale short-video and livestream platforms, model uncertainty can warn of low-quality recommendations that may lead to disengagement of LAUs and at the same time identify opportunities to diversify content recommendation for HAUs. To leverage this dichotomy, we introduce a unified, production-ready framework that calibrates uncertainty to drive differentiated strategies. Specifically, we implement a model-uncertainty-based risk-averse deboosting policy for LAUs to suppress unreliable recommendations, while employing a risk-seeking Upper Confidence Bound (UCB) strategy for HAUs to encourage exploration. Validated on a major livestream platform, our framework demonstrates significant improvements in retention (active hours) and satisfaction (quality watch time ratio) for LAUs as well as remarkable increases in interest diversity and category coverage for HAUs, proving the value of uncertainty-aware recommendation in industrial settings.

\end{abstract}



\maketitle

\section{Introduction}


Modern recommender systems face distinct uncertainty challenges across user segments. For Low-Active Users (LAUs), sparse interaction history makes inference fragile, where high uncertainty risks immediate churn. Conversely, High-Active Users (HAUs) possess abundant data, allowing uncertainty to serve as a signal for proactively expanding interest diversity. We argue that epistemic uncertainty should be treated as a risk for LAUs but as an exploration opportunity for HAUs. To this end, we propose a dual strategy: stabilizing LAU experiences through risk-averse deboosting, while encouraging HAU exploration via an Upper Confidence Bound (UCB) mechanism.

Unfortunately, most modern large-scale recommendation models output only point estimates with no inherent notion of confidence or uncertainty. Recent empirical analyzes confirm that state-of-the-art deep learning models can be poorly calibrated in practice — a high predicted score does not necessarily imply a correct or reliable recommendation \cite{penha2021calibration, angelopoulos2023recommendation}. This lack of an actionable uncertainty signal makes it difficult for ranking systems to distinguish "easy" predictions from "risky" ones. In high-traffic production environments, unrecognized model uncertainty can lead to significant user experience degradation by taking high-confidence actions on erroneous predictions for vulnerable user segments. These issues highlight the operational need for robust uncertainty-aware recommendation systems that not only predict user preferences, but also quantify—and act upon—the reliability of those predictions.

While the need for uncertainty quantification is clear, existing paradigms struggle to meet the dual requirements of low-latency inference and fine-grained granularity demanded by large-scale recommender systems. Traditionally, Bayesian approaches \cite{zhu2009risky, gopalan2014content} and their deep learning approximations like Monte Carlo (MC) dropout \cite{gal2016dropout} offer principled uncertainty estimates. However, they typically rely on expensive posterior sampling or multiple stochastic forward passes, introducing latency overheads that are prohibitive for high-concurrency real-time ranking. Conversely, distribution-free methods like Conformal Prediction \cite{angelopoulos2023recommendation} are computationally efficient but often produce global error guarantees (e.g., a single error rate averaged across all users). These "one-size-fits-all" intervals fail to capture input-specific risk—they cannot distinguish whether a specific video is risky for a specific user, often leaving relative rankings invariant and rendering them ineffective for personalized filtering.

This creates a critical operational gap: standard methods are either too slow (e.g., MC Dropout) or too coarse (e.g., Conformal Prediction) to handle the nuanced needs of diverse user segments. To address this, we propose a unified, production-ready framework that bypasses heavy sampling in favor of Input-Specific Expected Prediction Error (EPE) estimation. For standard point-estimate models, we train an auxiliary critic network to predict the expected error of each user-item pair, providing a fine-grained measure of epistemic uncertainty. For probabilistic architectures (e.g., Beta/Gamma outputs), we adopt an empirical Bayes procedure that infers input-dependent priors, enabling a principled decomposition of predictive variance into model-driven and data-driven components without sacrificing serving latency.


To translate these estimates into robust decision-making, we calibrate the uncertainty signals on a real-world hold-out dataset to establish reliable confidence thresholds. At serving time, our system applies a segment-aware dual strategy: for Low-Active Users (LAUs), high-uncertainty recommendations are rigorously deboosted to prevent churn caused by unreliable content; conversely, for High-Active Users (HAUs), the same signal serves as an exploration bonus to uncover new interests and expand diversity. This approach adapts uncertainty controls to distinct lifecycle stages, simultaneously strengthening short-term retention for vulnerable users while fostering the long-term vitality of the recommendation ecosystem.

Our work makes the following contributions, specifically within the context of industrial-scale recommendation systems:

\textbf{1. A Production-Ready, Unified Uncertainty Quantification Framework.}  
    We introduce a novel and flexible framework that augments existing recommender systems with input-specific uncertainty quantification capabilities. The framework unifies two practical approaches: an auxiliary critic network that learns to predict instance-specific Expected Prediction Error (EPE), and an empirical Bayes module designed for efficient variance estimation in probabilistic models. The combined uncertainty signals quantify confidence on a per-user, per-item basis, directly supporting safer, more robust ranking in real-world deployments.

\textbf{2. A Targeted Deboosting Policy for Low-Active Users (LAUs).}  
    We implement and deploy a practical uncertainty-aware deboosting strategy that leverages calibrated uncertainty thresholds to proactively filter or down-weight risky recommendations. This policy specifically addresses the cold-start and sparse-data challenges associated with LAUs on livestream platforms, mitigating the risk of user dissatisfaction and churn caused by unreliable predictions.

\textbf{3. An Uncertainty-Driven Exploration Strategy for High-Active Users (HAUs).}
    We propose a proactive exploration mechanism that utilizes high epistemic uncertainty as a positive signal for interest expansion. By applying an exploration bonus based on the Upper Confidence Bound (UCB) principle, the system prioritizes under-explored content domains for HAUs. This strategy effectively enhances recommendation diversity and facilitates content discovery by rewarding model-driven uncertainty, thereby encouraging the exploration of the user's latent interest spaces.

\textbf{4. Large-Scale Empirical Validation and Business Impact Analysis.}  
    Using large-scale production data from a major social media platform, we validate the framework's effectiveness. We demonstrate significant improvements in key business metrics for the vulnerable LAU segment: the deboosting policy substantially increases **user retention** (active hours) and downstream satisfaction (**quality watch time ratio**), while maintaining overall platform performance. These results underscore the operational value of uncertainty modeling in industrial recommendation systems and provide a generalizable strategy for improving reliability in practical, high-stakes environments.

\section{Methodology}
\label{sec:method}

We focus on two widely deployed model classes in real-world recommendation systems:  
(i)~\emph{point-estimation models}, such as deterministic CTR predictors and staytime regressors, and  
(ii)~\emph{probabilistic models}, which output parameters of a predictive distribution over the outcome.  
Our framework introduces two complementary uncertainty measures---one tailored to each model class---and integrates them into a unified strategy for user-tailored ranking.

\subsection{Definition of Uncertainty}

For each user-context pair input $x \in \mathcal{X}$, let $f_{\Theta}(x)$ denote the recommender’s prediction, where $\Theta$ represents the model parameters after daily training. As the model is continually fine-tuned using fresh data, $\Theta$ is naturally viewed as a random variable drawn from the implicit parameter distribution induced by the training pipeline.
Uncertainty for input $x$ arises from the following two sources.

\textbf{(1) model (epistemic) uncertainty}, including (i) offline model limitations, where restricted model capacity or limited training data lead to biased or unstable predictions; and (ii) online serving inconsistencies, where system-level factors such as feature staleness and serving latency cause deviations between offline predictions and real-time inference.

\textbf{(2) data (aleatoric) uncertainty}, arising from inherent randomness in the conditional outcome distribution $Y \mid X=x$.  

In the context of recommendation systems, we focus on estimating model uncertainty because
epistemic uncertainty directly reflects how well the system has learned the underlying user--item interaction patterns.  
High epistemic uncertainty reflects scenarios where the available behavioral signals are still evolving, particularly for cold-start users or segments with sparse interactions. In these cases, our framework optimizes for prediction robustness to ensure a smooth transition as the model's knowledge base matures. 
In such cases, the model's lack of confidence should inform downstream ranking and deboosting decisions, since unreliable recommendations can harm user experience and slow long-term learning.  
By isolating and quantifying epistemic uncertainty, our framework allows the system to identify inputs where predictions are fragile, guide cautious recommendation, and allocate exposure toward regions where additional data can meaningfully improve the model.

\subsection{Uncertainty for Point Estimation: Expected Prediction Error}
\label{subsec:epe}

Point-estimation models output a deterministic value $f_{\Theta}(x)$ but do not characterize uncertainty.  
We therefore study the \emph{input-specific expected prediction error} (EPE):
\[
\mathrm{EPE}(x)
=\mathbb{E}_{\Theta}\,\mathbb{E}_{Y\mid X=x}\bigl[\ell\bigl(Y, f_{\Theta}(x)\bigr)\bigr],
\]
where $\ell(\cdot,\cdot)$ is any nonnegative task-appropriate error measure. When $\ell$ is chosen to be the squared loss, the expected prediction error admits a canonical variance--bias decomposition that links model uncertainty and data uncertainty to distinct statistical components (other Bregman losses admit analogous decompositions, but squared loss provides the clearest interpretation):
\begin{equation*}
\label{eq:epe-decomp}
\begin{split}
\mu(x) &:= \mathbb{E}[Y \mid X=x],
\qquad 
\bar f(x) := \mathbb{E}_{\Theta}\!\left[f_{\Theta}(x)\right].\\
\mathrm{EPE}(x) &= \underbrace{\bigl(\bar f(x)-\mu(x)\bigr)^{2}}_{\text{model bias}}
\;+\;
\underbrace{\mathbb{E}_{\Theta}\!\left[\bigl(f_{\Theta}(x)-\bar f(x)\bigr)^{2}\right]}_{\text{model variance}}
\;+\;
\underbrace{\mathbb{V}(Y\mid X=x)}_{\text{data noise}}.
\end{split}
\end{equation*}
This decomposition provides an interpretable decomposition of uncertainty in our setting.

\textbf{Model (epistemic) uncertainty.}The term 
    $\mathbb{E}_{\Theta}[(f_{\Theta}(x)-\bar f(x))^{2}]$ measures instability of predictions across random training realizations, while $(\bar f(x)-\mu(x))^{2}$ captures systematic mismatch between the model family and the true conditional mean.  
    Both arise from finite data, optimization randomness, or representational limitations of the model class. Large model variance signals insufficient model confidence at $x$.

\textbf{Data (aleatoric) uncertainty.}
    The irreducible variability $\mathbb{V}(Y\mid X=x)$ reflects inherent randomness in user behavior conditional on the same features $x$, and cannot be reduced even with a perfect model. Large data noise indicates that user responses are inherently difficult to predict, even under an optimal model. 

In practice, our point-estimation models do not output these components individually; instead, the scalar $\mathrm{EPE}(x)$ aggregates their combined magnitude. Consequently, if one wishes to study or compare a \emph{specific} component of uncertainty across different inputs $x$, it is essential to hold the other component approximately fixed---that is, to control for factors that influence it.  
In our application, our goal is to use $\mathrm{EPE}(x)$ as a proxy for how well the model has learned (i.e., to compare epistemic uncertainty across inputs), and as a result, \textbf{we will restrict comparisons to be made within strata where intrinsic behavioral randomness is believed to be similar (e.g. $\mathrm{EPE}(x)$ is compared only among low-active users whose aleatoric variability is presumed comparable).} Under this condition, we define our model uncertainty score for point prediction models as $U_{\mathrm{point}}(x)\equiv \mathrm{EPE}(x),$ given Var$(Y|x)$ is approximately constant.

\subsection{Learning a Critic to Predict EPE}
\label{subsec:critic}

Suppose daily training produces multiple daily model checkpoints $\Theta^{(1)},\dots,\Theta^{(K)}$ trained using daily samples $(x^{(k)}_j,y^{(k)}_j)$. We compute realized generalization errors
\[
e_j^{(k)} = \ell\bigl(y^{(k)}_j, f_{\Theta^{(k-1)}}(x^{(k)}_j)\bigr),
\]
which serve as Monte Carlo samples approximating $\mathrm{EPE}(x^{(k)}_j)$. Note that the error evaluation is performed using the latest model checkpoint that has not been trained on $k$-th day samples.

We train a separate \emph{critic model} $g_{\phi}:\mathcal{X}\to\mathbb{R}_{\ge 0}$ to estimate $\mathrm{EPE}(x)$ directly from input features and internal signals from the recommender.

\textbf{Multi-view input representation.}
Let $h_{\Theta}(x)$ denote the intermediate embedding produced by the recommender and $f_{\Theta}(x)$ its final output.  
We construct the critic input
\[
z_{\Theta}(x) = [\,x \,\Vert\, h_{\Theta}(x) \,\Vert\, f_{\Theta}(x)\,],
\]
which captures both external covariates and how the recommender internally represents the instance.

\textbf{Decoupled training.}
To prevent feedback loops, we freeze the recommender during critic training.  
The critic is trained daily on tuples $\{z_{\Theta^{k-1}}(x^{(k)}_j), e_j^{(k)}\}$ collected from previous-day checkpoints. The critic is optimized under the standard mean-squared-error regression objective, and the resulting point-estimation uncertainty score is $
U_{\mathrm{point}}(x) \equiv g_{\phi}(z_{\Theta}(x)).$

\subsection{Uncertainty for Probabilistic Models}
\label{subsec:prob-uncertainty}

Probabilistic models output parameters of a \emph{conditional prior} 
$p(\theta \mid \gamma(x)),$
whose hyperparameters $\gamma$ (e.g., shape, rate, scale, dispersion) are produced by the model as functions of the input $x$.  
Given this conditional prior and a likelihood model $p(y \mid \theta)$, the resulting predictive distribution is
\begin{equation}
\label{eq:pred-integral}
p(y \mid x)
\;=\;
\int p(y \mid \theta)\, p(\theta \mid x)\, d\theta,
\end{equation}
where the integral is taken with respect to the latent parameter $\theta$.  

The model parameters $\Theta$ (trained via maximum marginal likelihood / type-II MLE) specify how the prior depends on $x$.  
Once the model is fitted, the predictive distribution \eqref{eq:pred-integral} has a well-defined mean and variance that can be used as point prediction and uncertainty measure respectively:
\[
\mu(x) = \mathbb{E}[Y \mid X=x], 
\qquad 
U_{\mathrm{total}}(x)=\mathrm{Var}(Y \mid X=x).
\]

Because the prior is input-dependent, the predictive variance naturally splits into two components via the law of total variance:
\begin{align}
\mathrm{Var}(Y \mid X=x)
&=
\mathbb{E}_{\theta \mid x}\!\left[\mathrm{Var}(Y \mid \theta)\right]
\;+\;
\mathrm{Var}_{\theta \mid x}\!\left(\mathbb{E}[Y \mid \theta]\right).
\label{eq:total-var}
\end{align}
Again, one can interpret the two terms as:
\begin{itemize}
    \item \textbf{Data (aleatoric) uncertainty:}
    \(
    \mathbb{E}_{\theta \mid x}\!\left[\mathrm{Var}(Y \mid \theta)\right],
    \)
    capturing randomness inherent to the user’s behavioral outcome even under fixed parameters.
    \item \textbf{Model (epistemic) uncertainty:}
    \(
    \mathrm{Var}_{\theta \mid x}\!\left(\mathbb{E}[Y \mid \theta]\right),
    \)
    capturing uncertainty about the latent predictive parameter due to limited or noisy evidence at this input $x$.
\end{itemize}

This decomposition is particularly important in our recommendation setting.  
Epistemic uncertainty reflects insufficient information about user--item interactions; when it is high, the model has not yet reliably learned $f(x)$.  
Such cases warrant cautious recommendation, either because additional data can eventually resolve the uncertainty (e.g., cold-start scenarios) or because reliable learning may be fundamentally limited (e.g., low-active-user cases).  
In contrast, high aleatoric uncertainty indicates that user responses are inherently variable but still informative: recommending in these regions can promote exploration and improve estimates over time, since the system only learns after making recommendations.  
Motivated by this distinction, we define the probabilistic uncertainty score
\[
U_{\mathrm{prob}}(x)
\equiv
\mathrm{Var}_{\theta \mid x}\!\left(\mathbb{E}[Y \mid \theta, X=x]\right),
\]
which isolates the model-driven component of uncertainty.  
This $U_{\mathrm{prob}}(x)$ constitutes the second pillar of our uncertainty-aware ranking policy.

\subsection{Training of Probabilistic Models via Maximum Marginal Likelihood}
\label{subsec:mml-training}

The probabilistic framework described above is trained by maximizing the \emph{marginal likelihood} of the observed outcomes.  
For each input--label pair $(x_i, y_i)$, the model specifies an input-dependent prior $p(\theta \mid \gamma(x_i))$ and a likelihood $p(y_i \mid \theta)$.  
The resulting marginal likelihood for a single observation is
\[
p(y_i \mid x_i; \Theta)
=
\int p(y_i \mid \theta)\, p(\theta \mid \gamma_{\Theta}(x_i))\, d\theta,
\]
where $\Theta$ denotes all trainable parameters governing the mapping $x \mapsto \gamma_{\Theta}(x)$.
Model training proceeds by solving the type-II maximum likelihood problem:
\[
\Theta^{\ast}
=
\arg\max_{\Theta}
\sum_{i=1}^{n}
\log 
\int p(y_i \mid \theta)\, p(\theta \mid \gamma_{\Theta}(x_i))\, d\theta.
\]

This objective encourages the model to select input-dependent priors that best explain the observed data after integrating out the latent variable $\theta$.  
Importantly, no point estimate of $\theta$ is learned; instead, uncertainty in $\theta$ is preserved throughout training, enabling the predictive distribution to capture both data noise and epistemic uncertainty.

\subsection{Empirical Calibration of Uncertainty Scores}
\label{subsec:calibration}

To convert raw uncertainty magnitudes into actionable thresholds, we use an empirical calibration set $\mathcal{D}_{\mathrm{cal}}$.  
For each $(x_j,y_j)\in\mathcal{D}_{\mathrm{cal}}$, compute
$U_{\mathrm{point}}(x_j), U_{\mathrm{prob}}(x_j).$ For a given quantile level $q$ (e.g.\ $0.95$), define calibrated thresholds: $$
\tau_{\mathrm{point}} = \mathrm{Quantile}_{q}(U_{\mathrm{point}}),
\tau_{\mathrm{prob}}  = \mathrm{Quantile}_{q}(U_{\mathrm{prob}}).
$$
These thresholds serve as uncertainty tolerance levels reflecting real-world error distributions without requiring parametric assumptions.

\subsection{Uncertainty-Aware Deboosting for Low-Active Users}
\label{subsec:masking}

Let $r_{u,i} = f_{\Theta}(x_{u,i})$ be the base relevance score for user $u$ and item $i$.  
For each candidate item, compute both uncertainty signals:
$U_{\mathrm{point}}(x_{u,i}),  U_{\mathrm{prob}}(x_{u,i}).$ An item is considered \emph{risky} if either score exceeds its calibrated threshold:
\[
\mathbb{I}_{\mathrm{risky}}(u,i)
=
\mathbb{I}\!\left[
U_{\mathrm{point}}(x_{u,i}) > \tau_{\mathrm{point}}
\;\;\text{or}\;\;
U_{\mathrm{prob}}(x_{u,i})  > \tau_{\mathrm{prob}}
\right].\]
\noindent
We apply a deboost score $D>0$ to all risky items:
\[
r^{\mathrm{final}}_{u,i}
=
r_{u,i} - D\cdot\mathbb{I}_{\mathrm{risky}}(u,i).
\]
\noindent
Final ranking is obtained by sorting the deboosted scores:
\[
\mathrm{Rank}_{\mathrm{final}}(u)
=
\mathrm{argsort}\Bigl(\{\,r^{\mathrm{final}}_{u,i}\,:\, i\in\mathcal{I}(u)\}\Bigr).
\]
\noindent
The hyperparameters $D$ and $(\tau_{\mathrm{point}},\tau_{\mathrm{prob}})$ are tuned via online A/B testing.

\subsection{Uncertainty-Driven Exploration Boost for High-Active Users}
\label{subsec:masking}

For high-active users (HAUs), high epistemic uncertainty serves as a critical signal for identifying unexplored content domains, offering opportunities to enhance recommendation diversity without compromising session continuity. To foster systematic interest exploration, we follow the Upper Confidence Bound (UCB) principle, which advocates for "optimism in the face of uncertainty". We treat the calibrated uncertainty signals as an exploration bonus to elevate items with high discovery potential.

We define the total epistemic uncertainty for HAUs as $U_{\mathrm{HAU}}(x_{u,i}) = \max(U_{\mathrm{point}}, U_{\mathrm{prob}}).$ 
The boosted ranking score is calculated as: $r^{\mathrm{final}}_{u,i} = r_{u,i} + \omega \cdot U_{\mathrm{HAU}}(x_{u,i}),$
where $\omega > 0$ is the exploration weight adjusted via A/B testing. This mechanism encourages the system to prioritize items with high discovery potential and long-term utility, thereby proactively increasing categorical coverage and optimizing for long-term recommendation diversity.
\section{Related Work}

\subsection{Bayesian modeling of recommendation uncertainty}
Early Bayesian models, such as Poisson factorization, naturally quantify uncertainty and handle sparse feedback effectively by treating interactions as probabilistic draws \cite{gopalan2014content, gopalan2015scalable, wang2018confidence}. However, these approaches struggle with scalability due to expensive posterior inference or lack the expressivity of modern deep learning models. Subsequent efforts like Confidence-Aware Matrix Factorization (CMF) introduced variance terms to estimate reliability, but they largely retained the sampling complexity of fully Bayesian methods and often failed to capture fine-grained, pair-specific uncertainty \cite{knyazev2023lightweight}. In the deep learning era, Monte Carlo (MC) dropout emerged as a solution, using stochastic forward passes to approximate epistemic uncertainty. While effective at the instance level, the computational cost of repeated sampling makes it impractical for low-latency industrial pipelines \cite{gal2016dropout}.

Our work positions empirical Bayes as a pragmatic middle ground. Rather than sampling full posterior distributions, we estimate prior hyperparameters directly from the data. This approach captures key Bayesian benefits—such as stabilizing predictions for low-activity users via partial pooling—and yields per-instance uncertainty estimates, all while avoiding the heavy computational overhead of traditional inference.

\subsection{Distribution-free modeling of uncertainty: Calibration and Error Estimation}
Accurate uncertainty quantification is often pursued via model calibration, where techniques like temperature scaling \cite{guo2017calibration} or isotonic regression ensure predicted scores align with true engagement probabilities. While effective in large-scale production systems like YouTube and LinkedIn \cite{bai2023regression, borisyuk2024lirank}, these methods typically apply global or bucket-wise adjustments, addressing population-level reliability rather than distinguishing uncertainty for specific user-item pairs. Alternatively, distribution-free frameworks like conformal prediction provide rigorous statistical guarantees by constructing predictive sets with bounded error rates \cite{angelopoulos2023recommendation}. However, these approaches introduce operational complexity—such as variable-length recommendation lists—and rely on global thresholds that lack fine-grained discrimination.

In contrast, our approach utilizes a critic network to predict the expected prediction error (EPE) for each individual recommendation. This yields a granular, input-specific uncertainty score that is empirically calibrated and operationally actionable. Unlike complex interval predictors, our metric integrates seamlessly into the ranking stack as a direct deboosting factor, combining the rigor of calibration with the flexibility of instance-level risk assessment.

\subsection{Applications for Low-Activity and Cold-Start Users}

Handling users with limited interaction history remains a longstanding challenge \cite{ricci2011recommender, zawia2025comprehensive}. Industrial platforms typically employ hybrid approaches combining collaborative filtering with auxiliary content features \cite{l2012recommender}. Recent research has pivoted toward meta-learning and rapid adaptation methods, such as MeLU \cite{lee2019melu} and M2EU \cite{wu2023m2eu}, which utilize MAML-based optimization to infer preferences from few examples. In production, specialized architectures like Cold \& Warm Net \cite{yin2023cold} and the MAIF framework \cite{cao2025industrial} handle zero-shot scenarios through multi-task learning.
Our work treats cold-start as an uncertainty quantification task \cite{xiang2025harnessing, zhang2025cold}. Frameworks like CREU \cite{xiang2025harnessing} leverage epistemic uncertainty to improve cold-start performance. By integrating a calibrated critic network, our framework provides a safeguard against high-uncertainty recommendations, which are particularly critical for maintaining the retention of users within their first 1,000-video tenure. 

\section{Experiments}
\label{sec:experiments}

This section presents our experimental setup, data, offline trend analyses, online A/B testing results, and ablation studies. All experiments are conducted on production-scale recommendation traffic on a major livestream platform, focusing on low-active users for whom model uncertainty has the strongest impact on ranking quality.

\subsection{Experimental Setup}
\label{subsec:setup}

We adopt a \textbf{DCN-based architecture} as the primary CTR predictor $f_{\Theta}(x)$ due to its proven efficiency and representational power in large-scale systems. To quantify uncertainty for deterministic predictions, we introduce an auxiliary \textbf{EPE Critic Head} that shares the DCN feature representation. This regression head is trained via MSE to approximate the expected prediction error,
\[
    \widehat{\mathrm{EPE}}(x) \approx \mathbb{E}_{\Theta} \mathbb{E}_{Y|x}\!\left[ (Y - f_{\Theta}(x))^2 \right],
\]
following the procedure outlined in Section~\ref{subsec:epe}.

In parallel, a \textbf{Bayesian Prior Head} projects shared features into logits $(u(x), v(x))$ to parameterize a dynamic prior $\theta \sim \mathrm{Beta}\big(\alpha(x), \beta(x)\big)$, where $
    \alpha(x) = 1 + \mathrm{softplus}(u(x)), 
    \beta(x) = 1 + \mathrm{softplus}(v(x)).$
Here, epistemic uncertainty is derived explicitly as the prior variance:
\[
    U_{\text{prob}}(x) = \frac{\alpha(x)\beta(x)}{(\alpha(x)+\beta(x))^2\,(\alpha(x)+\beta(x)+1)}.
\]
Together, these heads provide \textbf{complementary uncertainty signals}: the EPE critic targets error in deterministic point estimates, while the Bayesian head captures distributional model uncertainty.

To validate these methods, we compare against two established baselines. First, we implement a \textbf{Multi-head Ensemble Baseline} where the model shares all underlying embedding tables and DCN cross-layers, branching only at the final two fully-connected layers into $H=10$ distinct heads, with epistemic uncertainty estimated as the inter-head variance. Second, we utilize a \textbf{MC Dropout Baseline} \cite{gal2016dropout} by enabling stochastic dropout masks ($p=0.1$) during inference. We perform $T=10$ forward passes for each input sample and quantify uncertainty as the variance of the resulting predictive distribution, serving as a non-parametric approximation for Bayesian inference.

\begin{table*}[t]
    \centering
    \setlength{\tabcolsep}{12pt}
    
    \begin{tabular}{lcccc}
        \toprule
        \textbf{Model} & \textbf{Pearson} (Linear Correlation) & \textbf{Spearman} (Rank Correlation) & \textbf{AURC / Base} ($\downarrow$) \\
        \midrule
        Critic-Predicted EPE & 0.3335 & 0.6695 & 0.0157 / 0.0271 \\
        Bayesian Model & 0.3733 & 0.6881 & 0.0162 / 0.0277 \\
        Multi-head Ensemble & 0.0346 & 0.2760 & 0.0205 / 0.0270 \\
        MC Dropout & 0.0469 & 0.1887 & 0.0219 / 0.0264 \\
        \bottomrule
    \end{tabular}
    
    \caption{Comparative benchmarking of uncertainty estimators.}
    \label{tab:uncertainty_comparison}
\end{table*}

\begin{figure*}[t]
  \centering
  \begin{subfigure}[t]{0.3\textwidth}
    \centering
    \includegraphics[width=\textwidth]{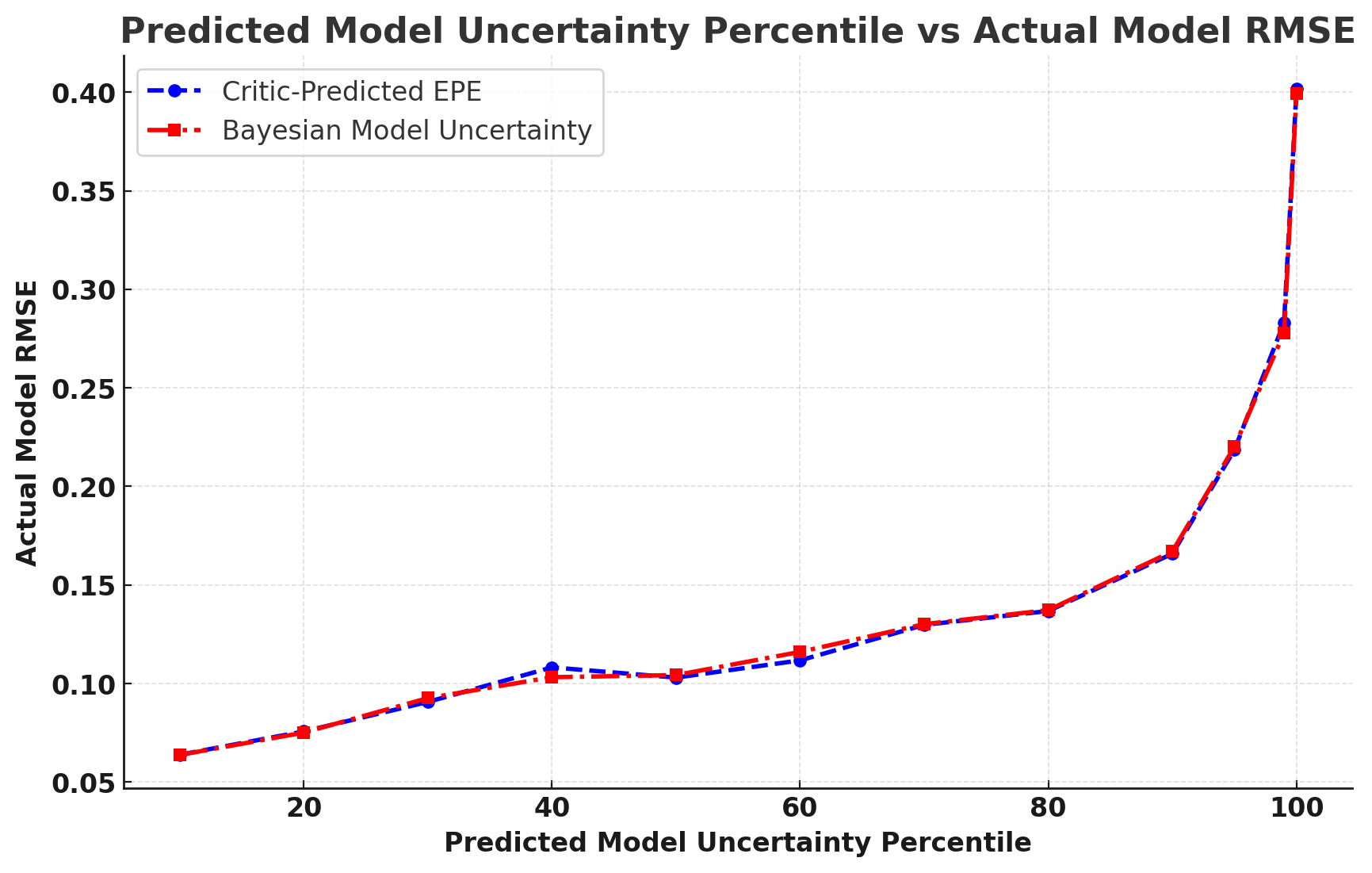}
    \caption{Realized Online MSE across predicted-uncertainty-percentile bins.}
    \label{fig:trend_epe_mse}
  \end{subfigure}\hspace{0.02\textwidth}
  \begin{subfigure}[t]{0.3\textwidth}
    \centering
    \includegraphics[width=\textwidth]{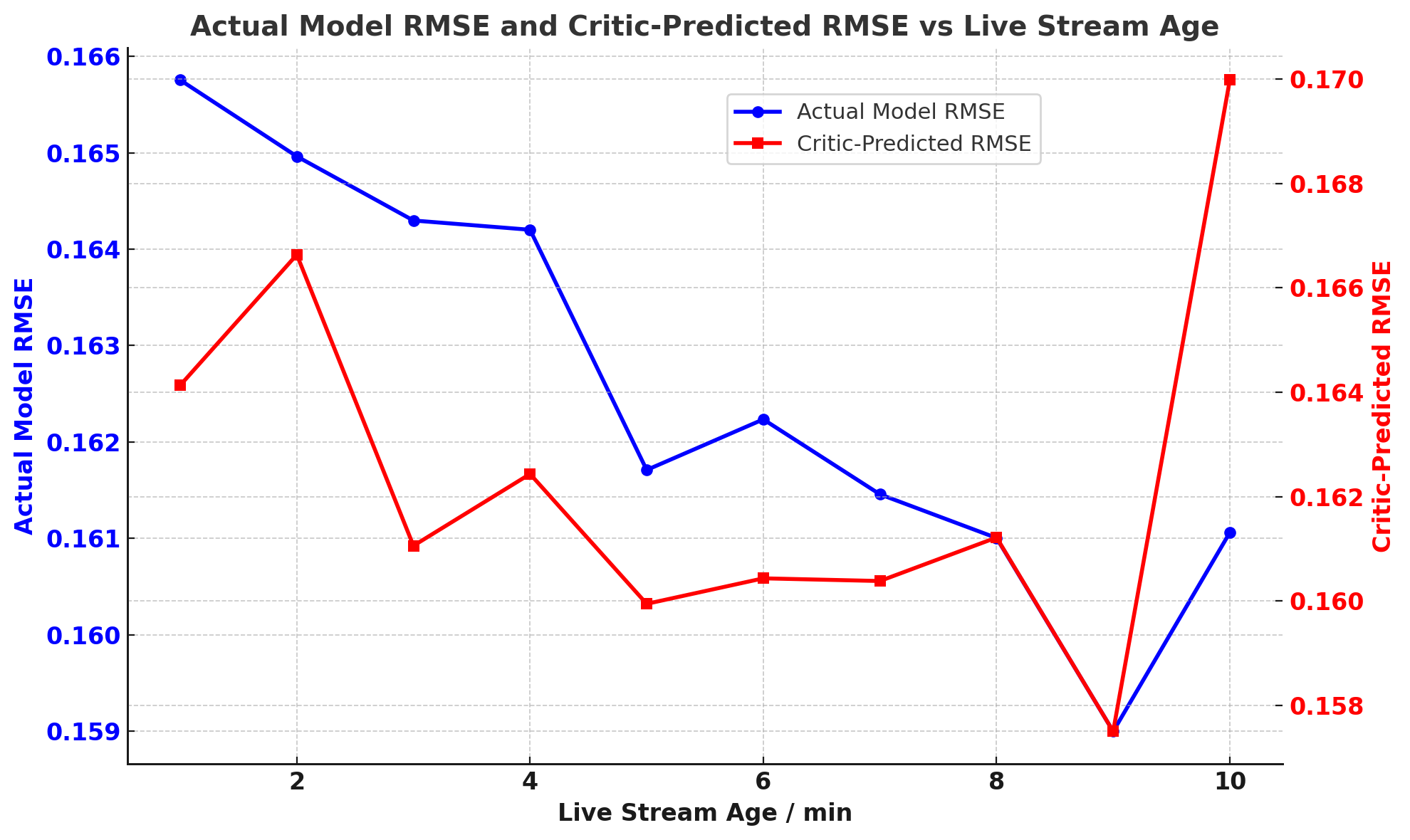}
    \caption{Critic-Predicted EPE uncertainty and realized RMSE versus live-stream age.}
    \label{fig:stream_age_epe}
  \end{subfigure}\hspace{0.02\textwidth}
  \begin{subfigure}[t]{0.3\textwidth}
    \centering
    \includegraphics[width=\textwidth]{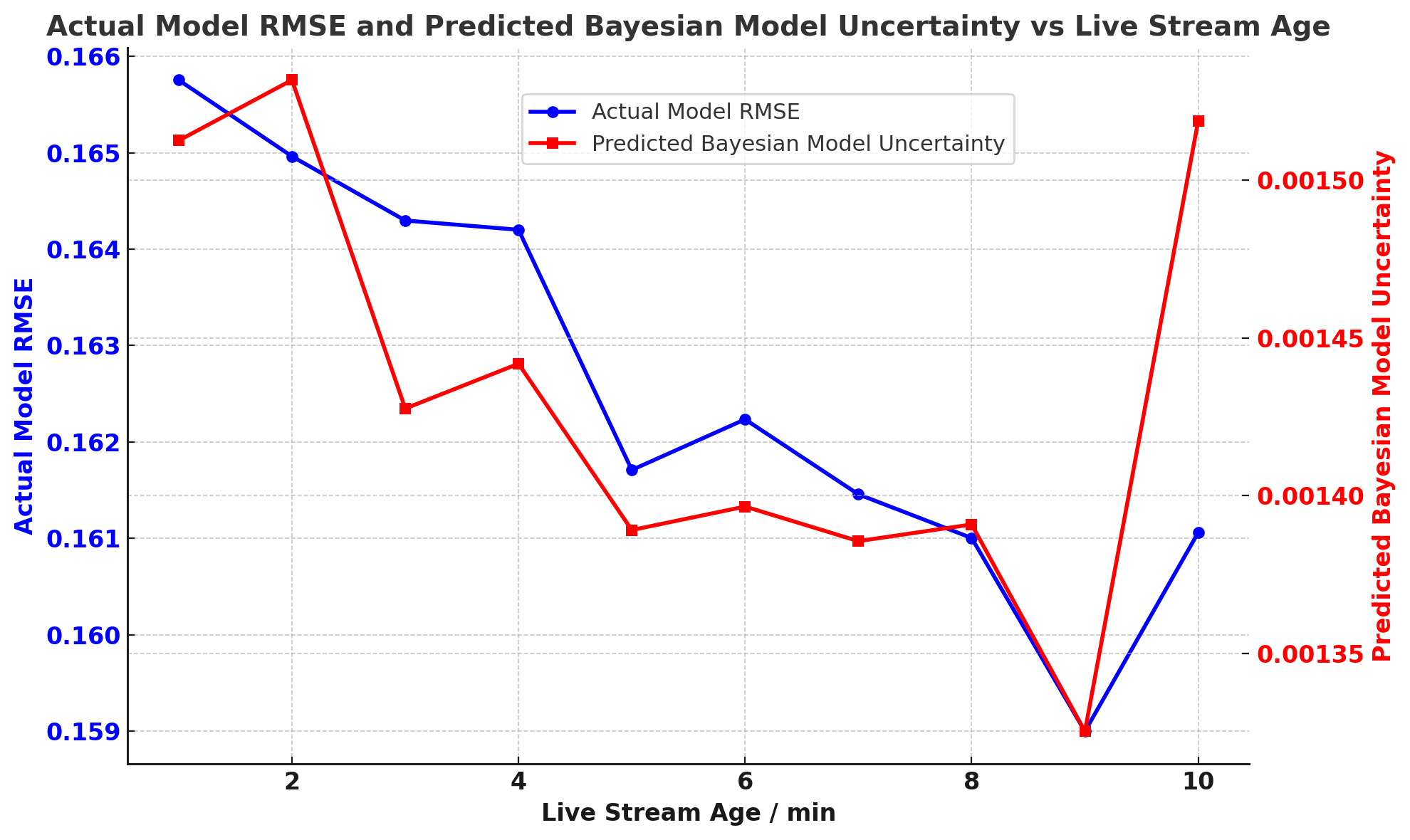}
    \caption{Bayesian model uncertainty and realized RMSE versus live-stream age.}
    \label{fig:stream_age_bayes}
  \end{subfigure}
  \caption{Trend analysis of uncertainty estimation.}
  \label{fig:three_panel_analysis}
\end{figure*}

\subsection{Data}
\label{subsec:data}

All models are trained and evaluated on anonymized production traffic from real users. 
We follow all internal data governance and compliance protocols. 
No personally identifiable information is used, and no cross-user identifiers are shared across training runs.

\subsection{Cohort Selection}
As mentioned in Section 2, we categorize users based on their historical engagement levels to ensure that aleatoric uncertainty is relatively consistent within each experimental group. In the livestream recommendation context, we define low-active users (LAUs) as those whose active watch days are less than 7 days in a month. Conversely, high-active users (HAUs) are defined as those whose active watch days are 7 days or more in a month. This segmentation allows us to evaluate the efficacy of our dual-strategy: applying risk-averse deboosting to stabilize experiences for LAUs, and leveraging UCB-based exploration to enhance content diversity for data-rich HAUs by proactively optimizing recommendation variety.

\subsection{Offline Analyses}
\label{subsec:offline}
Evaluating predictive uncertainty is challenging because ground-truth uncertainty is typically unobservable. In the literature, three primary paradigms are used for evaluation: 
(i) selective prediction metrics like Area Under the Risk-Coverage curve (AURC) \cite{geifman2017selective} to measure a model's efficiency in suppressing high-risk samples; (ii) correlation metrics including Spearman rank correlation to evaluate the monotonic relationship between uncertainty and error ranks \cite{kristoffersson2024uncertainty}, and the Pearson correlation coefficient to quantify the linear association between uncertainty estimates and observed prediction errors; and (iii) trend alignment diagnostics that verify consistency with domain-specific proxies like interaction density or content age.

\textbf{Uncertainty Performance.}
We evaluate the components of our unified framework—comprising both the EPE Critic Network and the Empirical Bayes (Bayes) prior head—against structural heuristic baselines using established quantitative metrics. Table 1 provides a comparative benchmark.

\textbf{Ranking Consistency.} Both Bayes and EPE Critic exhibit high Spearman correlations ($>0.65$), significantly outperforming ensemble heuristics. EPE Critic achieves the most competitive AURC / Base Risk ratio, confirming its superiority in risk-aware sample filtering. 



\begin{figure*}[t]
    \centering
    \begin{subfigure}[b]{0.48\textwidth}
        \centering
        \includegraphics[width=\linewidth]{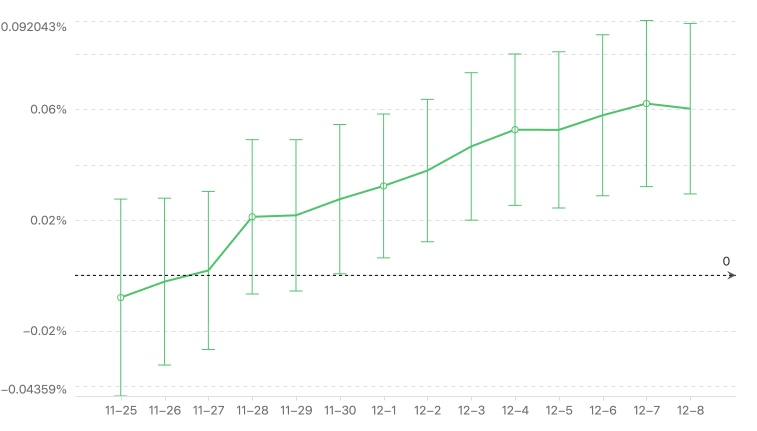}
        \caption{Improvements in HLT7}
        \label{fig:hlt_ab}
    \end{subfigure}
    \hfill
    \begin{subfigure}[b]{0.48\textwidth}
        \centering
        \includegraphics[width=\linewidth]{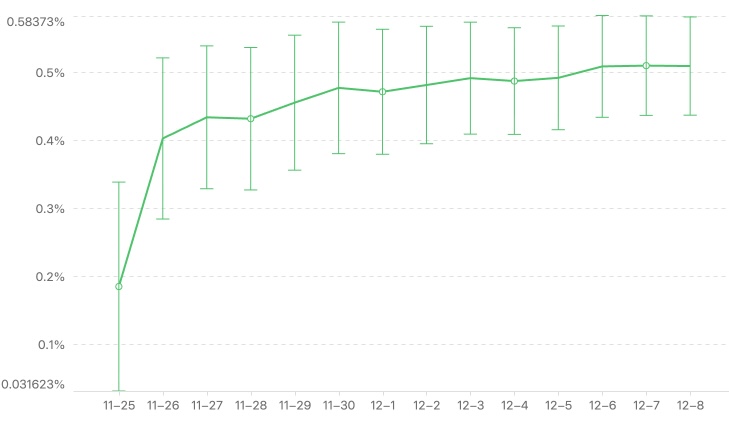}
        \caption{Improvements in valuable watch duration ratio}
        \label{fig:vwr_ab}
    \end{subfigure}
    \caption{A/B test results: daily-cumulative improvements over 14 days with 95\% confidence interval. (a) Results for HLT7. (b) Results for valuable watch duration ratio.}
    \label{fig:ab_test_comparison}
\end{figure*}

\textbf{Sanity Check I: Uncertainty vs.\ Realized Online MSE.}
Users are binned into decile buckets based on predicted uncertainty:
EPE-based $U_{\text{point}}(x)$ and Bayesian prior variance $U_{\text{prob}}(x)$.
For each bin, we compute the real-time RMSE of the online CTR model. As shown in Figure~\ref{fig:trend_epe_mse}, higher predicted uncertainty correlates with higher realized error. The relationship is monotonic across both uncertainty measures and highly consistent for both uncertainty measures.

\textbf{Sanity Check II: Uncertainty vs.\ Live-Stream Session Age.}
Live-stream recommendation provides a natural axis of uncertainty: a newly started stream has little behavioral history and is difficult to model, whereas a mature stream has accumulated evidence. We bin user--item requests by the live-stream age (minutes since stream start). 
Figures~\ref{fig:stream_age_epe} and~\ref{fig:stream_age_bayes} show that both uncertainty measures decrease with stream age, mirroring reductions in realized MSE.  
This validates that the uncertainty models behave consistently with domain intuition.

\subsection{Online A/B Testing}
\label{subsec:ab}

We deploy the uncertainty-aware strategy on production live-stream traffic for 14 days. The deployment strategy is cohort-specific: for Low-Active Users (LAUs), we apply a risk-aware deboosting policy; for High-Active Users (HAUs), we utilize high epistemic uncertainty as an exploration bonus (UCB principle) to stimulate interest expansion.

For the LAUs, we implemented an uncertainty-aware deboosting policy to mitigate the risk of undesirable recommendations arising from unreliable model predictions, which frequently trigger immediate disengagement and churn. To measure practical significance and business impact, we use a set core business and ecosystem indicators to capture the multi-dimensional impact of the framework: 

\noindent
\textbf{HLT7 (Hourly Life Time in Last 7 days)} Cumulative user active time, a standard indicator of sustained engagement and retention;

\noindent
\textbf{VWR (Valuable Watch Duration Ratio)} Ratio of valuable live watch time (organic engagement actions) to total live watch time;

\noindent
\textbf{App Stay Duration} Total daily time spent on the application, reflecting overall stickiness of the platform and potential spillover effects of live content quality on other application modules;

\noindent
\textbf{Live Watch Time} Raw duration of live stream consumption. In the context of our deboosting policy, we monitor this to ensure that improvements in engagement quality do not lead to an excessive contraction of total consumption volume;

\noindent
\textbf{HLT-efficiency} Defined as $-\Delta HLT7/\Delta \text{Live watch time}$. This composite metric quantifies the retention gain per unit of watch time sacrifice, measuring the efficiency of our policy in filtering out low-quality consumption to exchange for higher overall user lifetime.

Figures~\ref{fig:hlt_ab} and~\ref{fig:vwr_ab} show consistent and statistically significant improvements across all 14 days.
Filtering high-uncertainty items reduces detrimental recommendations for low-active users, leading to longer sessions and higher-quality watch behavior.

As summarized in Table~\ref{tab:ablation_results_lau_updated}, the unified framework achieved a +0.0577\% lift in HLT7 and a +0.512\% increase in VWR. The rise in App Stay Duration confirms a positive spillover effect, indicating that the safety-first strategy stabilizes the early-stage user journey. Crucially, while raw Live Watch Time showed a marginal decrease (-0.16\%) , the high HLT-efficiency (0.0549) demonstrates that the system successfully "pruned" high-risk, low-value content. This trade-off is scientifically honest: the platform sacrifices negligible passive duration to secure meaningful long-term retention gains.

For HAUs, we leverage model uncertainty as an exploration bonus to proactively enhance recommendation diversity. The goal was to prioritize recommendation diversity by optimizing the variety of content categories where the model's epistemic uncertainty is high. We evaluate the effectiveness using two diversity-centric metrics: 

\noindent
\textbf{Show Tag/U} Average number of unique livestream category tags shown per user, representing the breadth of topical exposure;

\noindent
\textbf{Top1 Tag UV Ratio} Proportion of users whose engagement within their leading interest domain exceeds 80\% of their total watch time; a decrease in this ratio reflects a continuous enhancement in recommendation diversity and a more balanced distribution across diverse content categories.

The UCB-based strategy (Table~\ref{tab:ablation_hau_final}) achieved a +2.1\% lift in Show Tag/U and a -0.42\% optimization of category distribution balance. These results confirm that the framework successfully increases recommendation diversity for HAUs while maintaining stable engagement, as evidenced by the non-significant increase in Live Watch Time (+0.16\%, $p > 0.05$).

\begin{table*}[htbp]
  \centering
  \begin{tabular}{lccccc}
    \toprule
    \textbf{Ablation Variant} & \textbf{HLT7} & \textbf{App Stayduration} & \textbf{Live Watch Time} & \textbf{HLT-efficiency} & \textbf{VWR} \\
    \midrule
    Unified Framework (EPE+Bayes) & +0.0577\% & +0.16\% & -1.05\% & 0.0549 & +0.512\% \\
    Only EPE Uncertainty         & +0.0434\% & +0.12\% & -1.07\% & 0.0404 & +0.08\% \\
    Only Bayesian Uncertainty    & +0.0468\% & +0.14\% & -0.95\% & 0.0491 & +0.3357\% \\
    Multi-head Ensemble          & +0.0188\%$^\dagger$  & +0.10\% & -0.81\% & 0.0232 & +0.0312\% \\
    MC Dropout                   & +0.0360\%$^\dagger$ & +0.07\% & -0.77\% & 0.0471 & +0.25\% \\
    Random Baseline              & +0.0029\%$^\dagger$ & +0.10\% & -2.38\% & 0.0012 & -0.0168\% \\
    \bottomrule
    \multicolumn{6}{l}{\small $^\dagger$ Not statistically significant.}
  \end{tabular}
  \caption{Ablation study results for LAUs. All numbers represent percentage lift relative to the production baseline. HLT7 metrics marked with $\dagger$ denote results that are not statistically significant ($p > 0.05$).}
  \label{tab:ablation_results_lau_updated}
\end{table*}

\begin{table*}[t]
  \centering
  \begin{tabular}{lccc}
    \toprule
    \textbf{Ablation Variant} & \textbf{Show Tag/U} & \textbf{Top1 Tag UV Ratio} & \textbf{Live Watch Time} \\
    \midrule
    Unified Framework (EPE+Bayes)         & +2.1\% & --0.42\% & +0.16\%$^\dagger$ \\
    Only EPE Uncertainty      & +1.0\% & --0.55\% & +0.07\%$^\dagger$ \\
    Only Bayesian Uncertainty & +0.9\% & --0.85\% & +0.03\%$^\dagger$ \\
    Multi-head Ensemble       & +0.82\% & --0.54\% & +0.09\%$^\dagger$ \\
    MC Dropout                & +0.7\% & --0.10\% & --0.20\%$^\dagger$ \\
    Random Baseline           & +0.16\% & +3.30\% & --1.62\%$^\dagger$ \\
    \bottomrule
    \multicolumn{4}{l}{\small $^\dagger$ Not statistically significant.}
  \end{tabular}
  \caption{Ablation study results for High-Active Users (HAUs). Show Tag/U and Top1 Tag UV Ratio measure interest expansion. Metrics marked with $^\dagger$ denote results that are not statistically significant ($p > 0.05$).}
  \label{tab:ablation_hau_final}
\end{table*}

\subsection{Ablation Studies}
\label{subsec:ablation}

To better understand how each component of our uncertainty-aware ranking strategy contributes to overall performance, we conduct a series of comprehensive ablation studies across both LAU and HAU traffic segments. We focus on two factors: (i) which uncertainty signal is used, and (ii) the magnitude of the recommendation strategy applied to high-uncertainty items.

\textbf{Ablation A: Using Only EPE Uncertainty.}
In this setting, deboosting and exploration decisions rely solely on the critic-predicted EPE $U_{\text{point}}(x)$, with the Bayesian model uncertainty removed. This variant tests whether the critic alone can support a robust ranking strategy. We observe +0.0434\% improvements in HLT7, but only +0.08\% gains in valuable watch ratio, indicating that EPE captures meaningful but incomplete model uncertainty.

\textbf{Ablation B: Using Only Bayesian Prior Variance.}
Here, deboosting and exploration decisions use only the Bayesian epistemic uncertainty $U_{\text{prob}}(x)$ computed from the Beta prior. This ablation evaluates whether uncertainty derived from latent parameter distributions is sufficiently informative on its own. Results show +0.0468\% gains in HLT7 and +0.3357\% in VWR for LAUs. Compared to the full model, these results suggest that Bayesian variance is complementary but less discriminative than the critic for immediate risk mitigation, while serving as a more stable driver for HAUs interest discovery (+0.9\% Show Tag/U). 

\textbf{Ablation C: Multi-head Ensemble Baseline.}
We replace our unified framework with a parameter-efficient multi-head ensemble. This variant tests whether structural disagreement alone is sufficient. We observe a minor +0.0188\% HLT7 lift and a +0.0312\% VWR lift for LAUs, alongside a +0.82\% lift in Show Tag/U for HAUs. These limited gains confirm that modal homogenization in shared representations yields uninformative uncertainty signals for complex DCN models, failing to provide the discriminative resolution needed for effective ranking interventions.

\textbf{Ablation D: Monte Carlo (MC) Dropout Baseline.}
We quantify uncertainty via 10 forward passes with stochastic dropout. This serves as a non-parametric variational approximation baseline. We observe a +0.0360\% improvement in HLT7 and a +0.25\% lift in VWR for LAUs. For HAUs, the Show Tag/U lift is only +0.7\%. Online performance is significantly lower than our proposed methods, indicating that random weight perturbations cannot capture instance-level epistemic risk as effectively as explicit modeling. 

\textbf{Ablation E: Random Baseline.}
To ensure that observed improvements are attributable to meaningful uncertainty estimation rather than incidental filtering, we construct a random baseline. Rather than relying on learned uncertainty estimators, this baseline generates a uniformly distributed random value $U_{rand} $ to serve as a pseudo-uncertainty score for each user-item pair. This variant yields neutral or slightly negative changes (+0.0029\% in HLT7, --0.0168\% in VWR) and fails to drive meaningful discovery (+0.16\% in Show Tag/U, +3.3\% in Top1 Tag UV Ratio), confirming that the benefits of our approach come from informed uncertainty signals rather than the act of deboosting/boosting itself.

\textbf{Summary of Ablation Results.}
Table~\ref{tab:ablation_results_lau_updated} and Table~\ref{tab:ablation_hau_final} summarizes the performance differences across all ablations. The results reveal that the Unified Framework (EPE + Bayesian) provides the most robust improvements, achieving the highest gains in both user retention (+0.0577\% HLT7) and interest discovery (+2.1\% Show Tag/U). By contrast, structural heuristics like Multi-head and MC Dropout exhibit diminished efficacy, and the Random Baseline results confirm that performance gains are strictly driven by the quality of informed uncertainty quantification rather than the ranking interventions themselves.

\textbf{Key Takeaways.}
Across all configurations, the Unified Framework (Full)—combining both explicit EPE prediction and Bayesian prior modeling—achieves the most robust performance gains. The Random Baseline confirms that observed improvements are strictly attributable to the quality of uncertainty quantification rather than incidental ranking perturbations. Notably, the significant performance gap between our approach and the Multi-head and MC Dropout benchmarks demonstrates that explicit density and error modeling are essential in industrial-scale architectures; traditional structural heuristics suffer from representation collapse, rendering them unable to capture the fine-grained epistemic risk necessary for effective risk mitigation and interest discovery.



\section{Discussions and Limitations}

In summary, our study demonstrates that incorporating explicit uncertainty estimation into large-scale recommendation systems can substantially improve reliability and long-term engagement across the user lifecycle. By combining critic-predicted Expected Prediction Error (EPE) with Bayesian prior variance, the system gains the "self-awareness" necessary to identify fragile predictions. Applying calibrated deboosting for Low-Active Users (LAUs) effectively mitigates the risk of disengagement caused by data sparsity, while UCB-based exploration for High-Active Users (HAUs) enhances recommendation diversity and promotes long-term ecosystem vitality. These differentiated strategies together yielded consistent gains in HLT7 (+0.057\%), valuable watch ratio (+0.512\%), and category discovery (+2.1\%).   

While effective in production, our approach has several limitations: it relies on stable daily retraining pipelines for generating EPE samples; the Beta prior used for CTR modeling, though computationally efficient, may oversimplify real-world behavioral variability in complex multi-modal environments; and the evaluation remains indirect as epistemic uncertainty itself lacks ground-truth labels. Furthermore, deboosting introduces inherent trade-offs between short-term safety and the long-term need for data acquisition that warrant deeper investigation. Notably, our current framework does not explicitly differentiate between heterogeneous sub-segments within the LAU cohort and may be sensitive to sudden distribution shifts in highly volatile livestream sessions.   

Future work may explore richer Bayesian neural networks based on Graph Neural Networks or Transformers to capture more complex interactions, unified frameworks that more tightly couple EPE and probabilistic components, and adaptive deboosting schedules using reinforcement learning. Integrating Large Language Models (LLMs) to provide semantic-aware uncertainty or explanations for exploratory content remains a promising frontier for enhancing user trust. Overall, our results highlight that uncertainty-aware ranking is a practical and impactful direction for building safer, more transparent, and more trustworthy recommendation ecosystems at an industrial scale.

\bibliographystyle{ACM-Reference-Format}
\bibliography{draft}

\end{document}